\documentclass[seceq]{ptptex}

\usepackage{graphicx}
\usepackage{wrapft}

\title{
FUSION03, Concluding Remarks%
}

\author{
A. B.  \textsc{BALANTEKIN}\footnote{E-mail: baha@nucth.physics.wisc.edu.}

}

\inst{
University of Wisconsin, Department of Physics, Madison, Wisconsin 
53706 USA 
}

\abst{
Fusion reactions below the Coulomb barrier provide new insights into
multidimensional quantum tunneling, nuclear reaction dynamics and
nuclear structure. These reactions are also of considerable interest to nuclear
astrophysics. In this summary recent developments in the
field are reviewed and open questions related to subbarrier fusion are 
presented.
}

\begin{document}

\maketitle

\section{Introduction}

Fusion reactions below the Coulomb barrier provide new insights into
multidimensional quantum tunneling, nuclear reaction dynamics and
nuclear structure \cite{Balantekin:1997yh,dg:ann}.
These reactions are also of considerable interest to nuclear
astrophysics. The evolution of main sequence stars, in particular
stellar nucleosynthesis is governed by subbarrier fusion
\cite{Adelberger:1998qm}.

\begin{wrapfigure}{l}{6.6cm}   
      \centerline{\includegraphics[width=6 cm,height=5 cm]{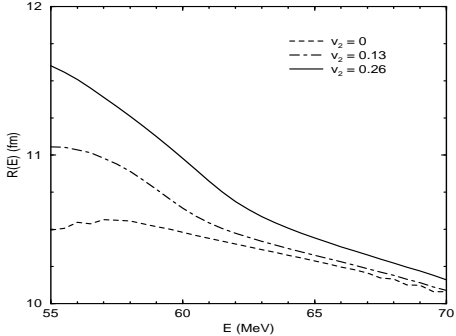}}
      \caption{$R(E)$ as a function of energy (see text).} 
      \label{fig:1}
    \end{wrapfigure}

It may be worthwhile to review some of the basic aspects of fusion 
reactions near and below the Coulomb barrier. The experimental 
observables are the cross section
\begin{equation}
\sigma (E) = \sum_{\ell=0}^{\infty} \sigma_{\ell} (E),
\label{eq1}
 \end{equation}
and the average angular momenta
\begin{equation}
\left\langle \ell (E) \right\rangle = \frac{{\sum_{\ell=0}^{\infty} \ell
\sigma_{\ell} (E)}}{\sum_{\ell=0}^{\infty} \sigma_{\ell} (E)}.
\label{eq2}
\end{equation}
The partial-wave cross sections are given by
\begin{equation}
\sigma_{\ell} (E) = \frac{\pi \hbar^2}{2 \mu E} (2 \ell +1) T_{\ell} (E),
\label{eq3}
\end{equation}
where $T_{\ell} (E)$ is the quantum-mechanical transmission
probability through the potential barrier and $\mu$ is the reduced
mass of the projectile and target system. 
For a one-dimensional barrier transmission probabilities
can be evaluated using a uniform WKB approximation
\begin{equation}
\label{eq4}
T_{\ell}(E) = \left[ 1 + \exp 
\sqrt{\frac{2\mu}{\hbar^2}} \int_{r_{1\ell}}^{r_{2\ell}} dr
\left[ V_0(r) + { \frac{\hbar^2 \ell(\ell+1)}{2\mu r^2}} - E \right]^{1/2}
 \right]^{-1} ,
\end{equation}
Under certain conditions \cite{Balantekin:1983dw}\tocite{Balantekin:1996dj} 
we can approximate the $\ell$ dependence of the transmission probability at 
a given energy by simply shifting the energy:
\begin{equation}
T_{\ell} \simeq T_0 \left[ E - \frac{\ell (\ell +1) \hbar^2}{2 \mu
R^2(E)} \right],
\label{eq5}
\end{equation}
where $\mu R^2(E)$ characterizes an effective moment of inertia.
In Ref. ~\citen{Balantekin:1996dj} it was shown that for a one-dimensional 
barrier $R(E)$ is a slowly varying function of energy (see Fig. 
\ref{fig:1}). Using Eq. (\ref{eq5}) and replacing summation in Eq. 
(\ref{eq1}) with an integration 
one obtains
\begin{equation}
E \sigma(E) = \pi R^2(E) \int_{0}^E dE' T_0(E').
\label{eq6}
\end{equation}
It is worth emphasizing that an $R(E)$ can be introduced not only for  
one-dimensional barrier penetration, but also to approximate the 
coupled-channels calculations. This is illustrated in Fig. \ref{fig:1}). 
In this figure the 
lowest (dashed) line corresponds to $R(E)$ extracted from a one-dimensional 
calculation for the system $^{16}$O + $^{154}$Sm. \cite{Balantekin:1996dj} 
The upper (dot-dashed and solid) lines illustrate $R(E)$ extracted from 
coupled channels calculations for the same system with the solid curve 
corresponding to a coupling strength twice a much as the strength used in 
calculating the dot-dashed curve. 
Eq. (\ref{eq5}) was inverted in Ref. ~\citen{Balantekin:1983dw} to obtain 
equivalent one-dimensional barriers. The inconsistency of those results 
indicated the necessity of a multidimensional coupled-channels picture 
for fusion reactions below the Coulomb barrier. 
 
Classically the tunneling probability is given by 
\begin{equation}
T_0(E) = \left\{ \begin{matrix} 1 & E \geq V_B \cr 0 & E<V_B \end{matrix}
\right. \>\>\> ,
\label{eq7}
\end{equation}
where $V_B$ is the barrier height. Hence replacing $R(E)$ in Eq. 
(\ref{eq6}) with a constant, usually taken to be position of the barrier 
height $r_0$, one obtains
\begin{equation}
E \sigma(E)  = \left\{ \begin{matrix} \pi r_0^2 (E-V_B) & E 
\geq V_B \cr 0 & E<V_B \end{matrix} \right. \>\>\> .
\label{eq8}
\end{equation}
Eq. (\ref{eq8}) was widely used to describe above the barrier fusion in 
the 1970's. From Eq. (\ref{eq8}) one can calculate the second derivative
\begin{equation}
\frac{d^2}{dE^2} [E \sigma(E)]  = \pi r_0^2 \delta (E-V_B) . 
\label{eq9}
\end{equation}
Quantum mechanically
this sharp peak is broadened as the transmission probability smoothly
changes from zero at energies far below the barrier to unity at
energies far above the barrier. When a number of channels are included 
the quantity in Eq. (\ref{eq9}) can be interpreted as the barrier 
distribution \cite{Rowley:1990gv} with the proviso that the energy 
dependence of $R(E)$ is not dominant. 

\section{Coupled-Channel Calculations}

The multidimensional barrier penetration problem inherent in subbarrier 
fusion needs to be addressed either in the coupled-channels \cite{dasso1} 
or the physically equivalent 
path-integral approach \cite{Balantekin:1984jv}. For numerical 
calculations coupled-channels formulation is the preferred method. In the 
earlier period of the study of subbarrier fusion simplified coupled-channel 
codes such as CCFUS \cite{dasso2}, CCDEF \cite{ccdef}, and CCMOD 
\cite{ccmod} were widely used. 
The precision of the current data requires a more accurate 
treatment of the channel-coupling problem going well beyond the 
approximations utilized in these codes. Currently newer codes are available 
to experimentalists, such as CCFULL \cite{Hagino:1999xb}, which try to 
eliminate at least some of these approximations. Erroneous 
interpretations of the data still appear in the literature. Most of these 
originate from using much older schemes, for example using a parabola to 
approximate the nuclear plus centrifugal potential. I would urge the 
experimental groups to analyze their data using the most precise 
coupled-channels  code available at the moment of their analysis.  

Various aspects of coupled-channels calculations were discussed in this 
conference by a number of speakers. Esbensen presented an analysis of 
the fusion of $^{27}$Al on a series of germanium isotopes. 
\cite{Esbensen:2003eg} He found that one needs to utilize nonlinear 
couplings for a reasonable description of the data, consistent with earlier 
results of other authors. \cite{Balantekin:1993gy,Hagino:1997kn}
In the channel coupling 
problem one usually assumes that at low energies nuclei start from their 
ground states. Takigawa discussed the implications of starting from an 
excited state. \cite{Kimura:2002vf} He showed that a significant number of 
transitions from the excited to the ground state may take place during fusion. 
A long-standing question in the coupled-channel calculations is to separate 
the effects of the surface excitations from those of the particle transfer 
reactions. Sensitivity to surface properties 
were explored by several speakers. Takigawa investigated the effects of 
double-folding potentials on the nuclear surface. \cite{taki1} Pollarolo, who 
analyzed the data for the fusion of $^{40}$Ca \cite{montagnoli} and 
$^{48}$Ca \cite{scaras} with Zr isotopes  also found that tunneling is 
dominated by the surface modes. \cite{Pollarolo} The consequences of taking 
different deformation parameters for neutrons and protons in coupled-channels 
calculations was discussed by Takigawa. \cite{taki1,taki2}

\section{The Nuclear Potential}

\begin{wrapfigure}{l}{6.6cm}   
      \centerline{\includegraphics[width=6.5 cm,height=4 cm]{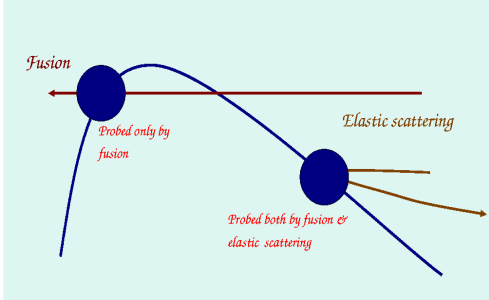}} 
      \caption{$R(E)$ as a function of energy (see text).}  
      \label{fig:2} 
    \end{wrapfigure}

A longstanding puzzle in the study of subbarrier fusion is the large values 
of the surface diffuseness parameter in the nuclear potential required to 
fit the data. In fact the value of the diffuseness one needs to fit the fusion 
data is typically 1.5 to 2 times the value of the diffuseness required to fit 
the elastic scattering data.\cite{Hagino:2001jj,Hagino:2003hj} One should 
emphasize that elastic scattering and fusion at subbarrier energies are 
complementary processes. 
While the outer part of the potential is probed both by elastic scattering and 
fusion processes, the inner part effects only the fusion (See Fig 
\ref{fig:2}). The inner turning point of the potential is much more sensitive 
to the value of the diffuseness parameter. Hence it may not be entirely 
surprising that fusion and elastic scattering can be fitted with somewhat 
different potentials. 

In this conference Dasgupta raised the possibility that 
this large value of the surface diffuseness may be a result of the 
dissipative  effects ignored in the coupled-channels codes.\cite{nanda} 
One should emphasize that calculations using an entirely different, 
algebraic, approach where nuclear structure effects are described by the 
interacting boson model and barrier penetration is calculated within a 
group-theoretical formalism also requires a large value of the surface 
diffuseness parameter.\cite{Balantekin:1993gz,Balantekin:1997bh} 
Very steep fall-off of the fusion data for the system $^{60}$Ni+$^{89}$Y 
at extreme subbarrier energies, described in this conference\cite{jiang}, 
is hard to understand using the canonical value of the 
diffuseness\cite{Jiang:2004zw}, but may be explained with a larger 
diffuseness parameter.\cite{Hagino:2003hj}  Such a large diffuseness parameter 
yields a shallow potential pocket, truncating the fusion cross section at 
energies below the minimum of the potential pocket when incoming-wave 
boundary conditions are used. 

\section{Inhibition of Fusion}

Another set of interesting results concerning fusion dynamics presented in 
this conference was studies of the fusion inhibition. Theoretically, when the 
product of the electric charges of the colliding nuclei exceed about 
$Z_1Z_2 \sim 1600$, 
one expects\cite{swia} that the large electrostatic energy of the system 
results in quasi-fission, inhibiting the fusion process. Experimental data, 
however, indicates that fusion is inhibited even when  $Z_1Z_2$ is much 
smaller. By studying three systems that lead to the compound nucleus 
$^{216}$Ra, namely $^{12}$C + $^{204}$Pb, and $^{19}$F + $^{197}$Au, and 
$^{30}$Si + $^{186}$W, the Australian National University group 
found\cite{nature} model-independent evidence for both quasi-fission and 
reduced fusion. Similarly fusion suppression and the presence of quasi-fission 
was observed in the systems $^{48}$Ca + $^{168,170}$Er\cite{Sagaidak:2003ph} 
and $^{48}$Ca + $^{154}$Sm\cite{trotta} at the Legnaro Laboratory. The latter 
groups find that the effect is more pronounced for more symmetric 
combinations and the target deformation may play a role in quasi-fission. 
In this conference a group from JAERI also reported fusion hindrance for 
the system $^{82}$Se + $^{138}$Ba\cite{jaeri}.     
It is worth emphasizing that while the fusion of asymmetric stiff systems is 
well-understood, symmetric systems present major theoretical challenges. 
For asymmetric systems the electromagnetic interaction is weaker, hence it 
takes the very tail of the nuclear potential to turn it around to form the 
potential barrier. The system tunnels through this barrier before nuclei get 
very close, hence a coupled-channels description in the basis of the 
truncated nuclear levels using incoming-wave boundary conditions is a good 
description of the underlying physics. In contrast, for heavy symmetric 
systems the barrier may be where the nuclei start touching each other, a 
truncated (as opposed to complete) nuclear level basis is inappropriate; other 
degrees of freedom such as transfer may come in. 

\section{Fusion of Unstable Nuclei}

The study of the fusion of unstable nuclei is a promising endeavor with an 
increasing number of new radioactive beam facilities coming on line. 
Preliminary efforts along this direction include a study of the coupling of 
the translational motion of the nuclei to a resonance 
state\cite{Balantekin:1984jv} and investigating the possibility of a 
molecular bond formation in the (experimentally difficult to achieve) system 
of $^9$Li + $^{11}$Li.\cite{Bertulani:1993hc} Key issues in the fusion of 
halo nuclei are separating complete and incomplete fusion and understanding 
the effects of transfer processes and positive Q-values on the barrier 
penetration. For such nuclei one needs to understand better the effect 
of the break-up process on the fusion cross section. There are a number of 
puzzles in the experimental data presented in this conference. For 
example $^{11,9}$Be + $^{209}$Bi data 
reported\cite{signor} are similar to the $^{10}$Be data 
even though $^{11}$Be is a halo nucleus. Similarly there seems to be exist 
a stripping break-up mechanism for the $^6$Li + $^{208}$Pb 
system.\cite{Signorini:2003yp,Wu:2003wp} 
On the other hand a very large subbarrier 
fusion enhancement was observed for the $^{132}$Sn + $^{64}$Ni 
system.\cite{Liang:2003hc} 
Hagino showed that\cite{Hagino:2004jk} although the continuum-continuum 
couplings reduce fusion above the barrier as compared to the no break-up 
case, they do not significantly change at below the barrier energies.  
Clearly much theoretical work needs to be done, in particular coupling to 
resonances should be better understood.

\section{Electron Screening}

Laboratory studies of subbarrier fusion of light nuclei are needed as 
input into the calculation of dynamics and evolution various 
astrophysical objects. Experimental methods are constantly improving: it 
is now possible to measure the fusion rates at Gamow peak energies in
underground laboratories with a very low background.\cite{Adelberger:1998qm} 
One current puzzle in these experiments is associated with electron 
screening; at very low energies appropriate for stellar conditions 
the atomic electrons of the fusing nuclei screen the Coulombic 
contribution to the potential. This contribution is usually calculated 
in the adiabatic approximation where one obtains a constant energy shift 
to the Coulomb potential.\cite{langanke} However the experimental data 
seem to require even larger shifts than those calculated in the 
adiabatic approximation. This is a long-standing puzzle as the adiabatic 
approximation is typically thought to overpredict. This discrepancy 
suggests some physics is excluded from the calculations. So far attempts 
to locate this missing physics have failed. Effects such as vacuum 
polarization, relativity, bremsstrahlung, and atomic polarization are 
very small.\cite{Balantekin:1997jh} It was suggested that virtual 
photon emission during tunneling may increase penetration 
probability.\cite{Flambaum:1998uu} The radiation field can be 
integrated out in the path-integral formalism; it turns out that this 
enhancement is extremely small.\cite{Hagino:2002pc} Any enhancement 
of the probability due to the break-up of the nuclei is also too 
small.\cite{Hagino:2003aq} In this conference Kimura showed that the 
screening potential may exhibit a radial variation in the tunneling 
region\cite{Kimura:2002ht}, which may increase the screening energy 
to beyond the adiabatic limit. Unfortunately this is not a 
satisfactory solution since this result depend on the excited state 
components. Discrepancy in screening continues to be a puzzle. 
In this conference Bertulani\cite{Bertulani:2004rv} 
explored if the stopping power is 
correctly extrapolated. 
Itahashi explored if experiments can be reconfigured to be free of 
screening corrections. 
In this regard the so-called Trojan horse method may help. This is a 
procedure to extract the astrophysical S-factor for two-body reactions 
by studying a closely-related three-body reaction under quasi-free 
scattering conditions. After an outline of this promising 
technique by Baur\cite{Baur:2004vw}  
successful applications of this method to the $^6$Li(p,$\alpha$)$^3$He 
reaction\cite{tumino}
and p-p elastic scattering\cite{pelleg} 
was discussed in this conference. 

\section{Fusion in Astrophysical Settings}

Nuclear fusion reactions play a very important role in astrophysical 
settings. Kajino\cite{Kaji} discussed  
nucleosynthesis in a core-collapse 
supernova and presented beautiful r-process nuclei abundance data 
obtained with the SUBARU telescope. 
Fiorentini\cite{fio} 
stressed that the Sun can be used as a laboratory to do 
fundamental physics and presented limits obtained on the p-p fusion 
reaction at very low energies using helioseismological observations. 

Nuclear reactions in a medium may be very different than in the 
laboratory (in the vacuum). This was highlighted by 
Kasagi\cite{kasagi} who 
discussed low-energy nuclear reactions in metals. 
Along those lines Rolfs discussed a very interesting possibility.
It looks like some deuterated materials give very high 
screening potentials. This is reminiscient of the Debye screening 
and raises the possibility of metals being similar to plasmas 
in this regard. 

\section{Conclusions}

This conference was held in one of the three famous scenic locations 
of Japan: The beautiful 
Matsushima Bay. The well-known poet Basho, who was a 
contemporary of Newton, visited Matsushima during his trip 
to the north of Japan. Literary lore tells us that he was speechless 
when he encountered the bay dotted with many islands covered with 
pine trees. His traveling companion Sora, however, was able to compose this 
Haiku:
\begin{verse}
Matsushima ya\\
Tsuru ni mi wo kare\\
Hototogisu.
\end{verse}
\begin{flushright}
Sora
\end{flushright}
This roughly translates into English as 
\begin{verse}
At Matsushima\\
Borrow your plumes from cranes\\
O nightingales.\footnote{It is amusing to note that this haiku can 
be translated into Turkish keeping its meter intact:
\begin{verse}
Matsushima'da\\
Turnan\i n sorgucunu\\
Als\i n b\"ulb\"uller
\end{verse}   }
\end{verse} 
Nightingales' own plumes were not sufficiently magnificent to do 
justice to the beauty of Matsushima. One feels the same way looking 
at the beautiful data presented in this conference. Many techniques 
we use to analyze the data are very simple-minded; the quality of 
the present data demands a better treatment with at least with 
full coupled-channels calculations, 
preferably using more microscopic approaches whenever it is possible.

\section*{Acknowledgments} 
I would like thank the organizers of the FUSION03 Conference for their 
hospitality and K. Hagino for his comments on the manuscript.   
This work was supported in part by the U.S. National Science
Foundation Grant No.\ PHY-0244384 and in part by
the University of Wisconsin Research Committee with funds granted by
the Wisconsin Alumni Research Foundation.

%

\end{document}